**Possibility of Exciton Mediated Superconductivity in Nano-Sized Sn/Si Core-Shell Clusters: A Process Technology towards Heterogeneous Material in Nano-Scale**

Yuichiro Kurokawa, Takehiko Hihara, Ikuo Ichinose[1], and Kenji Sumiyama[2]

*Department of Frontier Materials, Graduate School of Engineering, Nagoya Institute of Technology, Nagoya 466-8555, Japan*

[1]*Applied Physics, Graduate School of Engineering, Nagoya Institute of Technology, Nagoya 466-8555 Japan*

[2]*Department of Science, Graduate School of Science and Engineering, Tokyo Denki University, Hatoyama, Saitama 350-0394, Japan*

We have produced Sn/Si core-shell cluster assemblies by a plasma-gas-condensation cluster beam deposition apparatus.  For the sample with Si content = 12 at.%, the temperature dependence of electrical resistivity exhibits a metallic behavior above 10K and the onset of superconducting transition below 6.1 K.  With decreasing temperature, the thermomagnetic curve for the sample with Si content = 8 at.% begins to decrease steadily toward negative value below 7.7 K, indicating the Meissner effect.  An increase in the transition temperature, $T_C$ is attributable to exciton-type superconductivity.

Assembling of clusters is a promising approach to produce nanostructure-controlled-materials because their sizes are comparable to the elemental units of material properties. Several experimental techniques have been applied to prepare nano-sized clusters.[1-2] A plasma-gas-condensation cluster beam deposition (PGCCD) apparatus,[3-6] which is a combination of sputtering vaporization and inert gas condensation, is one of the promising candidates for this purpose. It is versatile to vaporize transition metals and refractive elements, and able to control the cluster size by adjusting the sputter yield, the gas pressure, and the volume of the cluster growth region. Using the PGCCD apparatus, we have succeeded in producing Co- or Fe-core/Si-shell, Fe-core/Al-shell, and Co-core/CoO-shell clusters assemblies,[7-10] in which the core parts are thermally and chemically stable, semiconducting and tunneling conductivities are observed.

In this study, we have tried to make an exciton-type superconductor with the Sn/Si core-shell cluster assemblies. Ginzburg suggested that $T_C$ of exciton-type superconductors are higher than those of conventional phonon-type superconductors.[11,12] Superconducting granules embedded in dielectric and/or semiconducting matrices,[13] and a sandwich structure consisting of alternate layers of metal and semiconductor have been proposed to be exciton-type superconductors,[12] in which electron-pairing force, i.e., exciton attraction, is mediated by virtual exciton in the semiconductor layers[12]. Indeed, superconductivity has been observed in Au/Ge and Au/Si multilayer systems.[14,15] It is assumed there that the superconductivity is due to the metallic phase of semiconductor.[16,17] However, Ginzburg point out possibility which those multilayer films are exciton-type superconductors.[13]

In this work, we have synthesized Sn/Si core-shell cluster assemblies and investigated their structures and electrical properties. The Sn/Si core-shell cluster assemblies were first prepared by the PGCCD apparatus. The apparatus is composed of three main parts: a sputtering chamber, a cluster growth room, and a deposition chamber. Sn and Si atoms were vaporized by Ar ions sputtering of Sn and p-type Si targets which installed face to face in the sputtering chamber. This process ensures that all clusters have nano-sized junctions without oxide layer by phase separation probably because they are immiscible each other: there is no solid solution in the Sn-Si equilibrium phase diagram.[18] Figure 1 shows a model of core-shell cluster formation in the PGCCD.

Figure 2 shows a transmission electron microscope (TEM) image and electron diffraction (ED) pattern of Sn/Si core-shell clusters, whose average chemical composition was estimated to be 37 at.% Si by means of energy dispersive X-ray spectroscopy (EDX).

In this figure, Sn cores are covered with Si shells for Sn/Si core-shell clusters with the diameter from 5 to 30 nm.  The ED pattern of Sn/Si core-shell clusters indicates that the clusters have a β-Sn structure.  The lattice constant estimated by X-ray diffractometer (XRD) for β-Sn in Sn/Si core-shell clusters with the average chemical composition of 28 at.% Si is $a = 5.83$ and $c = 3.18$.  We believe that Sn cores do not form solid solution with Si, because the lattice constant is equal to bulk Sn ($a = 5.83$, $c = 3.18$).[19]  The ED pattern of Sn/Si core-shell clusters, on the other hand, does not indicate diamond structure of Si suggesting that Si shell has an amorphous structure.  The inset of Fig. 2 exhibits the result of nano-beam EDX analysis on the same Sn/Si core-shell clusters.  As shown here, the low contrast areas are Si-rich and the high contrast ones Sn-rich.

According to the TEM observations, the average thickness of the Si shells on the Sn cores does not vary when the Si content changes from 3 to 37 at.%.  It is plausible that the Si vapor condenses to the primary particles of about 2 nm in diameter.  They are assembled on the Sn cores to form the shell structure.  Since the thickness of Si shell is almost constant, the volume ratio will be one of the better ways to label the sample than the atomic ratio in order to understand the physical properties of the present samples.  The Sn/Si cluster with the average chemical composition of 37 at.% Si, for instance, corresponds to 30 vol% Si.

Temperature dependence of electrical resistance, $R$ was measured using a physical properties measurement system (Quantum Design). At room temperature (RT), the electrical resistivity of Sn/Si core-shell cluster assembled film with 9 vol% Si is $2.9 \times 10^{-5}$ $\Omega$ m which is lower than that of the Sn cluster assembled film ($6.2 \times 10^{-5}$ $\Omega$ m) prepared by the same PGCCD apparatus. This result suggests that the oxidation of Sn cluster is suppressed by coating their surface with Si.

Figure 3(a) shows the temperature dependence of $R$ below 10 K in the applied magnetic field $H = 0$, 0.3, and 2 T, for the Sn/Si cluster assembled film. The present film exhibits a superconducting transition at $T > 4$ K which is higher than $T_C$ of bulk Sn ($T_C = 3.7$ K). The temperature at which $R$ in $H = 0$ is deviating from $R$ in $H > 0$ is reasonably defined as an onset temperature of superconductivity $T_{CO}$. $T_{CO}$ of the Sn/Si cluster assembled film is estimated to be 6.1 K that is higher than $T_C$ of bulk Sn.

Magnetization, $M$ was measured using a superconducting quantum interference device (SQUID) magnetometer in a magnetic field $H = 0.01$ T after zero field cooling. Figure 3(b) shows the temperature dependence of $M$ for the Sn/Si cluster assembled film with 6 vol% Si. It exhibits a paramagnetic behavior down to 10 K. With decreasing temperature, however, it begins to decrease steadily toward negative value below 7.7 K. The rapid decrease of $M$ is ascribed to a superconducting transition because the resistivity also begins to reduce at around the same temperature.

It is noteworthy that the TEM observations indicated the Si shells around Sn cores were incomplete, when the Si volume ratio was lower than 16%. In this case, the Sn cores coalesce with neighbor ones and form the networks with metallic conductivity over the substrate. Figure 4 shows electrical resistivity, $\rho$ vs temperature, $T$ for Sn/Si clusters assembled film with 23 vol% Si. The present film exhibits semiconductor-type temperature dependence because the conductive passes of Sn networks are intercepted by Si shells owing to higher volume ratio of Si shells.[20] It can be possible that the Si shells form solid solutions with Sn atoms because the plasma-gas-condensation is a non-equilibrium process. Figure 4 indicates however that the Si shells are not metallic with Sn doping but still semiconductive behavior.

The effect of the lattice softening at the cluster surface is one of the possible origins in the increase of $T_C$.[21] In such nano-sized clusters, $T_C$ increases with decreasing the cluster size $d$. Moreover, it has been observed in Sn cluster.[22-24] Matsuo *et al.* estimated $T_C$ values of nano particles of Al, In, and Pb from the following formula derived by

McMillan:[21,25)]

$$T_C = \frac{\Theta_D}{1.45}\exp\left[\frac{1.04(1+\lambda)}{\lambda - \mu^*(1+0.62\lambda)}\right], \qquad (1)$$

where $\mu^*$ is the Coulomb pseudopotential, which is almost 0.1 for polyvalent metals. $\Theta_D$ is the Debye temperature, which can be proportional to the average phonon frequency $<\omega>$. $\lambda$ is given as

$$\lambda = N(0)<g^2>/M<\omega^2>, \qquad (2)$$

where $M$ is the ion mass and $N(0)<g^2>$ is an empirical quantity. When the number of nearest-neighbor atoms on the surface is decreased, the force constant will decrease, and accordingly the average phonon frequency becomes low. Using eq. (1), we estimate the $T_C$ to be 4.7 and 3.9 K for the Sn/Si core-shell clusters with 20 vol% Si having Sn-core of 7 and 44 nm in diameter whose values are estimated to be smallest and largest ones by TEM observation, respectively. These $T_C$ are lower than $T_{CO}$ observed in the present work.

In the Sn/Si core-shell cluster assembly, moreover, the lattice softening is suppressed by the Si shell. Accordingly, in the present sample, virtual exciton in Si shells can mediate the interaction between electrons in Sn cores at Sn/Si interfaces because no oxide layer at Sn/Si interfaces which formed by phase separation in vacuum. The present result also proposes a potential route to fabricate new functional materials: introduction of a nano-scale heterogeneous structure by assembling nano-clusters.

Acknowledgement

This work has been supported by a Grant-in-Aid for Scientific Research from the Ministry of Education, Culture, Sports, Science and Technology Japan.

**Figure captions**

**Figure 1** Schematic diagram of a model for Sn/Si core-shell cluster formation.

**Figure 2** (a) TEM image of Sn/Si core-shell clusters whose average chemical composition is 37 at.% Si corresponding to 30 vol% Si. The inset shows EDX analysis of the same sample. (b) Corresponding electron diffraction pattern. The arrowed rings indicate formation of the β-Sn structure.

**Figure 3** (a) Electrical resistance $R$ as a function of temperature, $T$ for the Sn/Si core-shell cluster assembly whose average chemical composition is 12 at.% Si corresponding to 9 vol% Si. $T = 6.1$ K is temperature which $R$ of Sn/Si core-shell cluster begins to decrease. The circles, squares and triangles indicate $R$ which measured in magnetic fields = 0, 0.3, and 2 T. (b) Magnetization $M$ as a function of temperature $T$ in magnetic field $H = 0.01$T after zero field cooling of the Sn/Si core-shell cluster assembly whose average chemical composition is 8 at.% Si corresponding to 6 vol% Si. $T = 7.7$K is temperature which $M$ of Sn/Si core-shell cluster begins to decrease.

**Figure 4** Electrical resisitivity, $\rho$ as a function of temperature, $T$ for Sn/Si clusters assembled film whose average chemical composition is 29 at.% Si corresponding to 23 vol% Si.

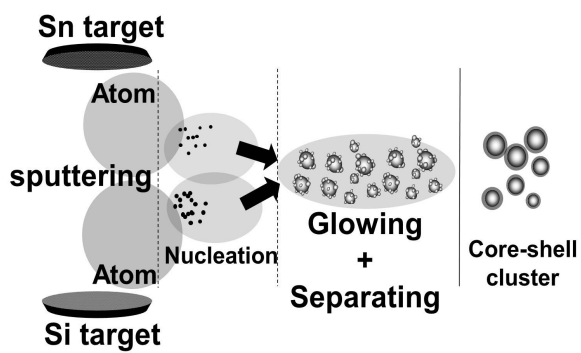

Fig. 1

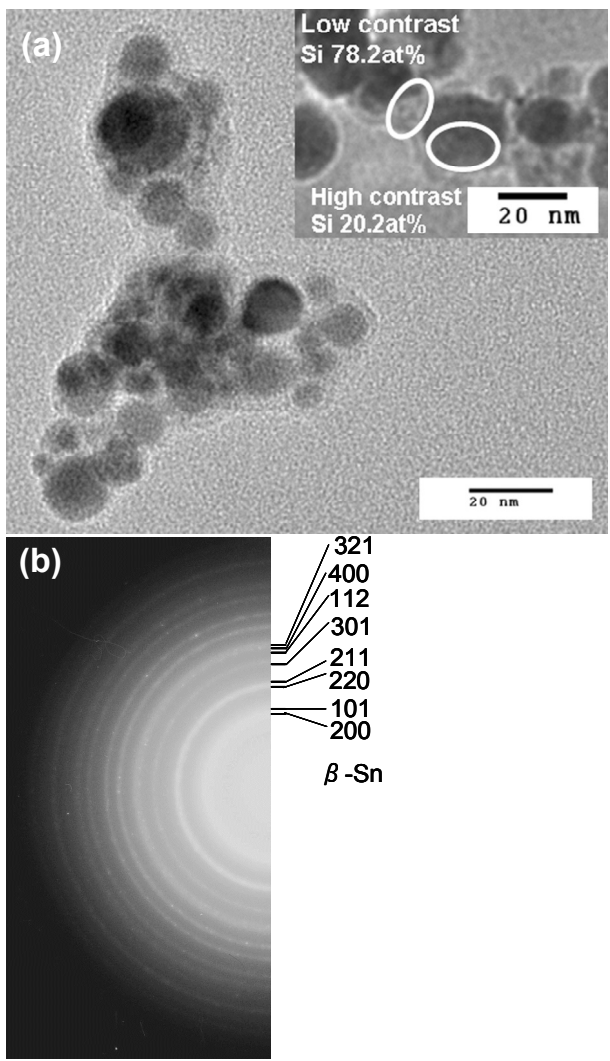

Fig. 2

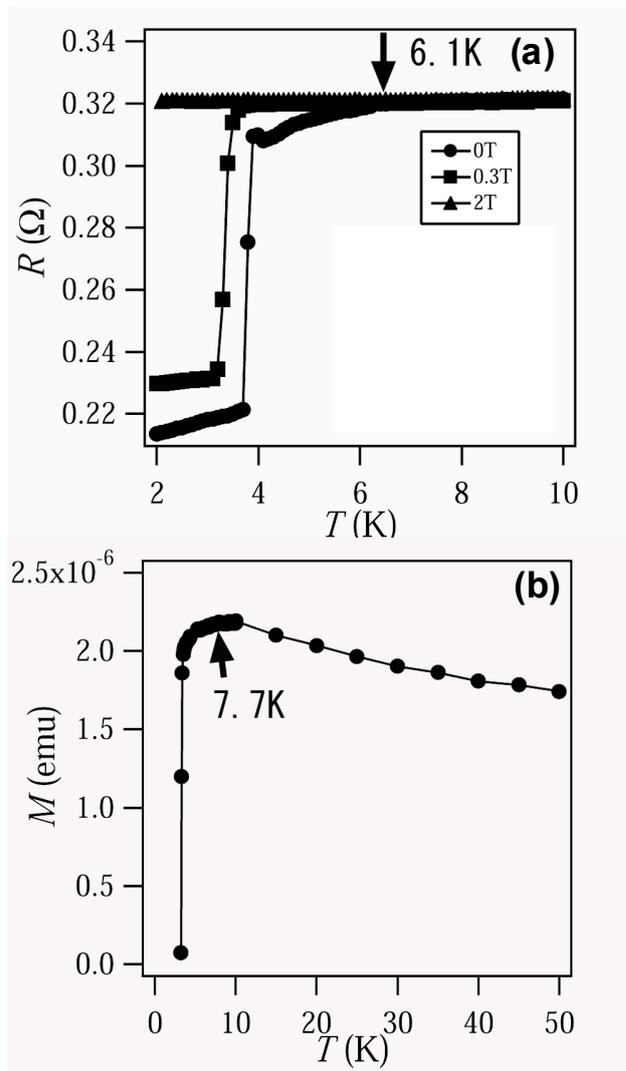

Fig. 3

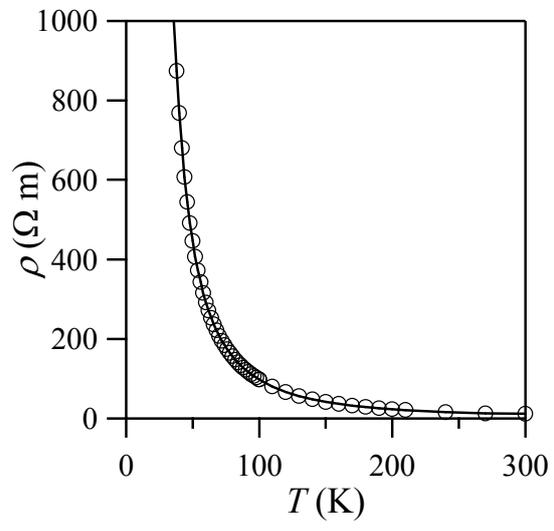

Fig. 4